\documentclass[preprint,showpacs,preprintnumbers,amsmath,amssymb]{revtex4}
\usepackage{graphicx}
\usepackage{dcolumn}
\usepackage{bm}
\begin{document}

\title{Limits on Nonstandard Forces in the Submicrometer Range}

\author{M. Masuda}
\email{masuda@icrr.u-tokyo.ac.jp}
\affiliation{Institute for Cosmic Ray Research, University of Tokyo, Chiba 277-8582, Japan}
\author{M. Sasaki}
\affiliation{Institute for Cosmic Ray Research, University of Tokyo, Chiba 277-8582, Japan}
\email{sasakim@icrr.u-tokyo.ac.jp}

\begin{abstract}
We report the search for a nonstandard force by measuring the Casimir forces in the 0.48--6.5~$\mu$m range. 
By comparing the data and the theory of the Casimir force, 
we have obtained constraints for the parameter $\alpha$ of the Yukawa-type deviations from Newtonian gravity. 
The obtained limits are more stringent than previous limits in the 1.0--2.9~$\mu$m range. 
Furthermore, we have obtained lower limits for the fundamental scale $M_{*}$ for gauged baryon number in the bulk. 
In particular, for six extra dimensions, the limits on $M_{*}$ are stringent in the range $6.5\times10^{-6}<\rho<2.5\times10^{-4}$. 
\end{abstract}


\maketitle
Quantum electrodynamics predicts that particle-antiparticle pairs will be continuously created and annihilated in vacuum.
This microscopic phenomenon results in a macroscopic force between conductors, 
called the Casimir force~\cite{Casimir}. 
We have focused on the fact that the existence of a macroscopic force 
that lies outside the standard model can produce deviations from the Casimir force.
Therefore, we developed a high-precision apparatus for detecting the anomaly~\cite{MasudaSasakiAraya}. 
A number of innovative theories that shatter the conventional concepts of time-space and fundamental interactions predict that 
nonstandard forces become obvious in the range in which the Casimir force is dominant~\cite{Dimopoulos0,Arkani2,Dimopoulos}.
In particular, the proposed apparatus is effective for demonstrating a number of models that assume extra dimensions. 
We can investigate such nonstandard forces that revolutionize fundamental physics by 
searching for deviations from standard quantum electrodynamics.

The Casimir force, as predicted by standard quantum electrodynamics, 
between a flat conductor and a spherical conductor is expressed by 
\begin{eqnarray}
\label{Cas}
F_{cas}(z)=\frac{{\pi}^3{\hbar}c}{360}\frac{R}{z^3} \quad  \mathrm{for} \quad z\ll R,
\end{eqnarray}
where $z$ is the distance between the conductors and $R$ is the radius of the spherical lens. 
When calculating the Casimir force between real metals, 
corrections for the properties of the metals 
(such as finite conductivity, nonzero temperature, and surface roughness) must be taken into account. Accurate computations can be performed for the finite conductivity and the roughness corrections (see the review paper~\cite{Bordag0}). However, four different models for the nonzero temperature correction have been proposed~\cite{Bordag0,Bostrom2,Svetovoy2,Genet,Klimchitskaya3}. 
Of these four models, two have been discounted based on experimental measurements~\cite{Decca0}. 
The discrepancy between the nonzero temperature corrections calculated
using the remaining two models is much smaller than the experimental accuracy.
Thus, the existence of a nonstandard force can be demonstrated by 
searching for deviations between the theoretical and experimental Casimir forces. 

During the last decade, various experimental demonstrations of the Casimir force have been reported. 
Since Lamoreaux's measurement by a torsion balance~\cite{Lamoreaux1}, devices~\cite{Mohideen1,Decca3,Decca2} such as 
atomic force microscopes or microelectromechanical systems 
have frequently been used to measure the Casimir force.
However, torsion balances have several advantages over other devices
for probing nonstandard forces. 
These include a high sensitivity to macroscopic forces, 
because a torsional period on the order of 1000~s is attainable, and 
the availability of materials with large radius of curvature, which generate large macroscopic forces.
 
On the other hand, special care must be taken to ensure that the distance between the conductors remains constant when using torsion balances. 
This is the case not only because torsion balances are sensitive to 
disturbances such as seismic motions and ground tilts, 
but also because distance fluctuations produce large systematic errors
as a result of the strong dependence of the Casimir force on distance. 
Various noises, including thermal noise and detection noise, 
have been reported in torsion balance experiments~\cite{Lambrecht}\cite{Lamoreaux0}.
However, there has never been a quantitative estimation of the extent to which seismic motions and ground tilts affect the measurement of a macroscopic force in the sub-micrometer range. Thus, we have developed a torsion balance to measure macroscopic forces in the sub-micrometer range and estimated the effects due to the seismic motions and tilt effects~\cite{MasudaSasakiAraya}. 

\begin{figure}
\includegraphics
[width=8.6cm]{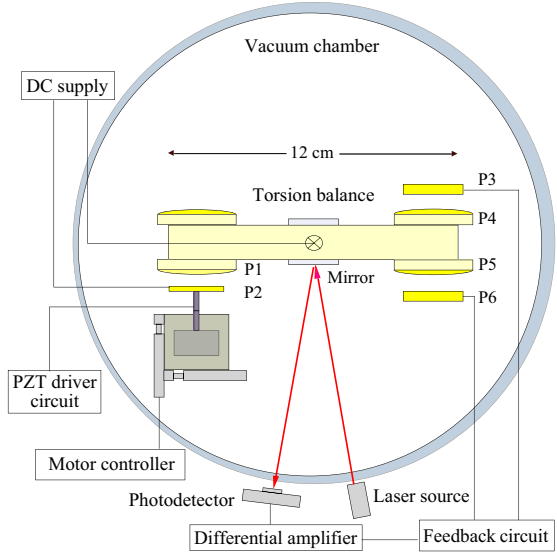}
\caption{Schematic diagram of the instrument used herein.  
P1 and P2 indicate the spherical lens and the flat plate, respectively, that produce a macroscopic force.
P3, P4, P5, and P6 indicate the plates forming actuators that are used to control the torsion angle. 
Plates P1, P4, and P5 are electrically grounded through the wire of the torsion balance.
}
\label{apparatus}
\end{figure}

 A schematic diagram of the present setup is shown in Fig.~\ref{apparatus} (details are shown in~\cite{MasudaSasakiAraya}). 
A spherical lens and a flat plate for generating the force are used to avoid the difficulty in the alignment of parallelism.
We use optical lenses made of BK7 glass as substrates for both plates. 
Spherical lens P1 has a radius of 20~mm, a thickness of 5~mm, a radius of curvature $R$ of 207~mm, and the root mean square amplitude of its surface roughness was measured to be 22~nm. 
Flat plate P2 has a radius of 15~mm, a thickness of 2~mm, and the root mean square amplitude of its surface roughness is nominally less than 10~nm.
These surfaces were metalized by evaporation coating a chromium layer with a thickness of 20~nm and then evaporation coating a gold layer with a thickness of 1000~nm. 

The torsion balance is a copper bar that is suspended above the center of gravity by a tungsten wire of diameter 60~$\mu$m and length 400~mm and the torsion constant of the wire is $6.1\times10^{-7}$~N$\cdot$m/rad.
The torsion angle can be obtained by an optical lever with a helium-neon laser and a photodiode.
The angular resolution of the photodiode was measured to be 2.3~mV/$\mu$rad, and 
the angular sensitivity was measured to be $1\times10^{-6}$~rad/$\sqrt{\mathrm{Hz}}$ at 1~mHz.

A feedback system is used to measure the macroscopic forces by a null method.
The torsion angle is maintained in order to maintain a certain position during the measurement. 
The signal of the torsion angle is integrated, filtered, and then converted into the feedback voltage ${\Delta}V$ in the analog feedback circuit. 
The position of the torsion balance is controlled by applying voltages $V_0+{\Delta}V$ and $V_0-{\Delta}V$ to plates P3 and P6 respectively, where $V_0$ is the bias voltage. 
The feedback response was adjusted to give a unity gain frequency of approximately 0.04 Hz. 
An electric force corresponding to ${\Delta}F=2V_{0}{\Delta}V dC/dx$ is applied to the torsion balance, where $dC/dx$ is the derivative of the capacitance of the plates. 
The feedback coefficient $\Delta F/\Delta V$ 
was calibrated to be $1.11\pm0.01 {\times} 10^{-12}$~N/mV. 

The variation of the force acting on plates P1 and P2 is determined by measuring the feedback voltage. 
Flat plate P2 is attached to the piezoelectric translator (PZT) on a motorized stage 
and faces the spherical plate P1. 
The dependence of displacement of the PZT on applied voltage (11.0~${\mu}$m at a bias voltage of 100~V ) was calibrated with a Michelson interferometer. 
The PZT was used only along the calibrated routes and 
the repeatability error in distance is measured to be 0.3 $\%$. 
These instruments were set in a vacuum chamber and 
a pressure on the order of 0.1~Pa was maintained during measurements.

In order to determine the absolute distance $z_0$ between plates P1 and P2, 
we use the dependence of the electric force on distance. 
If a bias voltage $V_{b}$ is applied between a flat plate and a spherical lens, 
the electrostatic force $F_e(V_{b})$ for $z\ll R$ is given by $F_{e}(V_b)={\pi}{\epsilon}_{0}R(V_{c}-V_{b})^{2}/z$,
where $V_{c}$ is the contact potential difference and ${\epsilon _0}$ is the dielectric constant.
The contact potential difference between two metallic surfaces is 
caused by the different work functions of different metals or the distribution of surface charges. 
If the bias voltage is varied with an amplitude of $V_s$ at a certain distance $z_0$, 
the variation in the electrostatic force $\Delta F_{e}(V_{b})\equiv F_{e}(V_{b}+V_{s})-F_{e}(V_{b})$ is given by 
\begin{eqnarray}
\Delta F_{e}(V_{b})=\frac{2V_{s}{\pi}{\epsilon}_{0}R}{z_0}V_{b}+\frac{{\pi}{\epsilon}_{0}R}{z_0}(V_{s}^{2}-2V_{c}V_{s}).
\label{elefit}
\end{eqnarray}

First, we measured $\Delta F_{e}(V_b)$ while the bias voltage was changed in steps of 10~mV at a certain distance. 
By fitting the data to Eq.~(\ref{elefit}), we obtained $V_c=82.6\pm0.9$~mV and $z_0=1.601\pm0.013$~${\mu}$m. 
We then measured the force as a function of distance by bringing plate P2 close to plate P1 
with a constant step of 0.3~${\mu}$m from a distance of approximately 6.5~$\mu$m. 
A bias voltage is applied between plates P1 and P2 to diminish the electric force. 
After obtaining data at a distance of approximately 0.5~$\mu$m, we returned the plate to approximately 6.5~$\mu$m 
and began the next cycle in the same manner. 
Each step requires 600~s, including a dead time of 200~s, 
and one cycle requires approximately 4 h. 
The final data comprises 587 points in the 0.48--6.5~$\mu$m range. 

\begin{figure}
\includegraphics
[width=8.6cm]{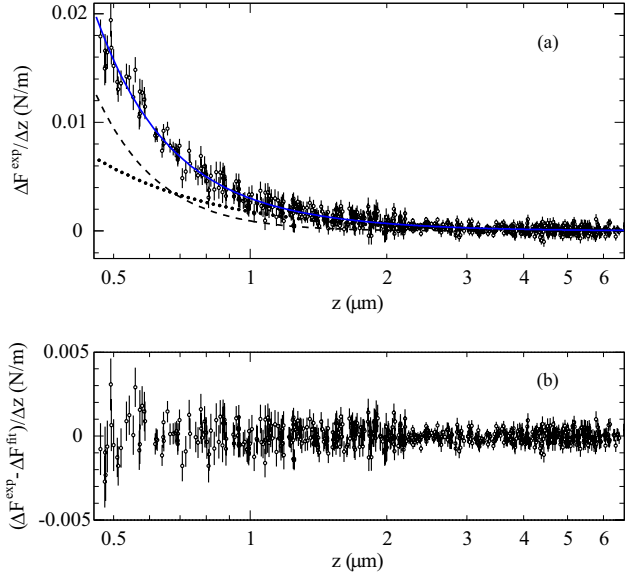}
\caption{(a) Data of the difference in force per constant step ${\Delta}F^{exp}(z)/{\Delta}z$ as a function of distance after qualification cut (See text). 
The vertical error bars show statistical errors. 
The dotted, dashed, and solid curves show the residual electric force,
the expected Casimir force after correction of the conductivity and roughness effects,
and the total best-fit curve of the sum of the two functions, respectively. 
(b)
Deviation between data of ${\Delta}F^{exp}(z)/\Delta z$ and 
the best-fit of the sum of the residual electric force and
the theoretical Casimir force
${\Delta}F^{fit}(z)/{\Delta}z$.
}
\label{fvsd}
\end{figure}

The obtained data correspond to the force variation ${\Delta}F^{exp}(z)\equiv F^{exp}(z)-F^{exp}(z+{\Delta}z)$, 
where $F^{exp}(z)$ is the attractive force between the plates as a function of distance. 
All data were divided into independent bins with the width of 0.1 to 0.3~${\mu}$m, fitted with a linear function defined within each bin.
Then, to ensure the quality of data analysis, 28 data points with 2.5~$\sigma$ far from the fitted function were rejected, 
which are 5$\%$ of the obtained data.
Within the scope of the standard interaction theory,
each data point conceals the residual electrostatic and the Casimir force variations.
The difference in the electric force when the distance is varied with a constant step 
${\Delta}z$ at a constant bias voltage can be expressed as 
\begin{equation} 
\label{denki}
\frac{{\Delta}F_{e}(z)}{{\Delta}z}
=\frac{F_{e}(z)-F_{e}(z+{\Delta}z)}{{\Delta}z}=\frac{{\pi}{\epsilon}_{0}RV_{res}^{2}}{z(z+{\Delta}z)}.
\end{equation}
In order to reveal the electric force and the Casimir force, 
the data of ${\Delta}F^{exp}(z)/{\Delta}z$ with the statistical errors 
in the whole range of the 0.48--6.5~$\mu$m
were fitted simultaneously to the sum of 2 functions:
the expected electrostatic force Eq.~(\ref{denki})
whose residual voltage difference $V_{res}$
is a fit parameter and the Casimir force 
without assuming any nonstandard interaction. 
The statistic error in distance is estimated from 
0.3$\%$ relative fluctuation of the movement of the PZT actuation. 
As a result, the residual voltage difference $V_{res}$ in Eq.~(\ref{denki})
is determined to be 20.0 ${\pm}$ 0.2~mV. 
The minimized value of $\chi^{2}/N_{dof}$ is obtained to be $513/558$.
The fairly reasonable goodness of fit qualifies the high level of 
agreement between the data and the standard interaction theory of Casimir force.
The data with the functions of the electric force (dotted), 
the Casimir force (dashed) 
with taking into account the conductivity and roughness effects, 
and the total best-fit curve (solid)
as the sum of 2 functions are as shown in Fig.~\ref{fvsd}(a).
The comparison between data and best-fit function is also shown in Fig.~\ref{fvsd}(b).
Experimental systematic errors were obtained by independently measuring the uncertainties in force and distance.
The uncertainty in force has a 1$\%$ relative error by measuring the feedback coefficient.
The uncertainty in distance of 13~nm is estimated from combining the offset error
and 0.3$\%$ relative fluctuation of the movement of the PZT actuation. 
$\chi^{2}/N_{dof}$ is slightly changed into 
$533/558$ and  $525/558$  with the same total function as obtained above,
as data are artificially shifted by $1\%$ in the force and 13~nm in the distance respectively.
Still the data fairly agree with the hypothesis of the total force composed of 
the residual electric and the standard Casimir force.

\begin{figure}
\includegraphics
[width=8.6cm]{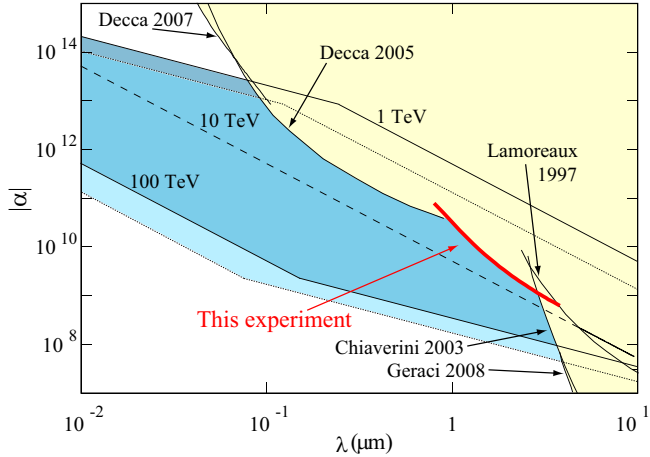}
\caption{Upper limits on the Yukawa parameter $\alpha$. These curves were obtained by experiments~\cite{Lamoreaux1}\cite{Decca3}\cite{Decca2}\cite{Chiaverini}\cite{Geraci}, respectively.
The curve labeled 'this experiment' shows the limits obtained at 95$\%$ CL. 
The shaded area enclosed by solid lines and the shaded area enclosed by dashed lines 
show the predicted regions for $\beta=1$ and $\beta=2$, respectively, for gauged baryon number in the bulk~\cite{Dimopoulos0}. 
The solid lines labeled 1~TeV and 100~TeV and the dashed line labeled 10~TeV
show the predicted line at the value of $M_{*}$ for $\beta=1$.  
}
\label{residual} 
\end{figure}
We then obtained the limits on nonstandard forces at around one micrometer. 
The potential of the nonstandard forces are generally expressed as the Yukawa-type potential, 
\begin{equation} 
\label{yukawa}
V_{Yu}(\lambda)=-\alpha G \int \mathrm{d}\mathbf{r}_1 \int \mathrm{d}
\mathbf{r}_2 \frac{\rho_1(\mathbf{r}_{1}) \rho_2(\mathbf{r}_{2})}{\mathbf{r}_{12}}\mathrm{e}^{-\mathbf{r}_{12}/\lambda },
\end{equation}
where $\rho_i$ for $i=1,2$ is the density of matter, $r_{12}$ is the distance between two bodies, $\alpha$ is the strength parameter; and $\lambda$ is the range parameter. 
Assuming that the geometry of this experiment is employed, 
the nonstandard force $F_{Yu}(z)$ is expressed from~\cite{Bordag2} as 
$F_{Yu}(z)=\alpha \sum_{i,j=1}^{3}\rho_{i}\rho_{j}f_{Yu}^{(i,j)}(z)$, 
where $\rho_{i}$ for $i=1,2$, and 3 is the density of BK7, chrome, and gold, respectively,  
and $f_{Yu}^{(i,j)}(z)$ are the parameters used in Eq.~(25)-(27) in~\cite{Bordag2}. 

We performed 2-parameter-fit between the data and the sum of 3 functions of the electric force, 
the Casimir force, and the Yukawa-type forces, simultaneously using 2 fit parameters of 
$V_{res}$ in Eq.~(\ref{denki}) and $\alpha$ in Eq.~(\ref{yukawa}) 
as changing the fixed parameter of $\lambda$ in Eq.~(\ref{yukawa}).
The 2$\sigma$ limit on $\alpha$ is obtained in each bin and then combined with the limit. 
In order to take into account systematic errors, 
limits on $\alpha$ are obtained by shifting data by 1 $\%$ in the force and 
13~nm in the distance, simultaneously. 
And then the most conservative 2$\sigma$ limits are shown in Fig.~\ref{residual}.  
Different limits of Lamoreaux's experiment are used in different papers (for example~\cite{Decca2} and~\cite{Long}). 
An overestimate of the accuracy of Lamoreaux's experiment was reported~\cite{Lambrecht}. In particular, the error in the distance was not considered. 
Thus, we used the conservative 1~$\sigma$ limits used in~\cite{Long} in Fig.~\ref{residual}. 
The other limits used in Fig.~\ref{residual} are at 95$\%$ CL. 
As a result, the limits on $\alpha$ obtained by this experiment are more stringent than previous limits in the 1.0--2.9~${\mu}$m range. 

\begin{figure}
\includegraphics
[width=8.6cm]{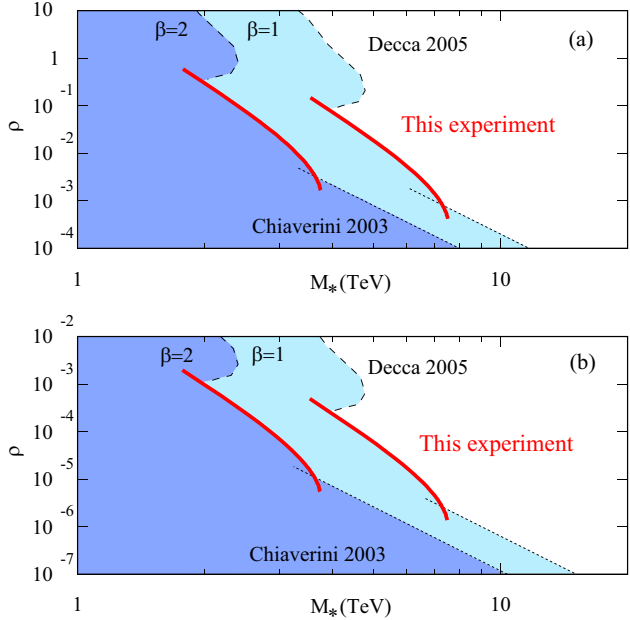}
\caption{Lower limits on fundamental scale $M_{*}$ at 95$\%$ CL for $n=3$ (a) and $n=6$ (b), respectively, 
assuming $\beta=1$ or $\beta=2$. 
The solid curves, the dashed curves, and the dotted curves are the present data, and the limits calculated from~\cite{Decca2} and~\cite{Chiaverini}, respectively.} 
\label{last}
\end{figure}
Using the limits on $\alpha$, we can search for nonstandard forces predicted by particle physics. 
In the theory of gauged baryon number in the bulk, it is predicted that the forces are more than five times stronger than gravity at less than 1~mm~\cite{Dimopoulos0}. 
There are two important parameters to describe the theoretical model, $\rho$ and $M_{*}$, where
$\rho$ represents the strength of gauge coupling in $n+4$ dimensions and 
$M_{*}$ is the fundamental energy scale, which is thought to be on the order of TeV~\cite{Dimopoulos0,Arkani2,Dimopoulos}. 
The other lower limits on $M_{*}$ have been indirectly obtained from astrophysical observations. 
As limits on the model of graviton in the bulk, 
based on the observation of PSR J0952+0755, it was reported that $M_{*}>1.61\times10^5$, $3.01\times10^2$, 25.5, 2.77 for one, two, three, and four extra dimensions, respectively~\cite{Hannestad}. It has been reported that the limits that apply for $n$ extra dimensions in the graviton case now apply to $n+2$ extra dimensions in the gauged baryon number in the bulk. 
However, it has also been reported that if there are fast decay channels for the  
Kaluza-Klein particles or if the dimensions are not of equal size, the constrains could be further weakened~\cite{Dimopoulos0}. 

 We have obtained experimental limits on the gauged baryon number in the bulk 
by measuring the Casimir forces in the sub-micrometer range. 
In particular, for $n=3$ and $6$, the obtained limits are as shown in Fig.~\ref{last}.  
In this calculation, it is assumed that ultraviolet cutoff scale $\Lambda$ =$M_{*}$,
as described in~\cite{Dimopoulos0}. 
These limits are obtained for $\beta=$1 and 2, where $\beta$ is defined as 
the ratio of vacuum expectation divided by $\Lambda $. 
As a result, in the case of $\beta=1$ and $n=6$, the obtained limits are stronger than previous limits in the range $6.5\times10^{-6}<\rho<2.5\times10^{-4}$. 

 We have demonstrated that we can probe the essence of particle physics by searching 
the vacuum, which means the ground state of energy, with a tabletop experiment. 
The experimental sensitivity of the present study is limited by seismic vibrations~\cite{MasudaSasakiAraya}, 
and we will explore unknown interactions more comprehensively in the sub-micrometer range not only by using an advanced attenuation system
but also by using high-precision displacement measurements with feedback frequency much higher than that of dominant seismic vibration.

We thank Professor A.\ Araya (University of Tokyo) and Professor Y.\ Higashi (KEK) for the use of their instruments and their useful technical suggestions. 
The present research was supported by KAKENHI (14047208 and 17654047) and cooperative use of the Mizusawa VERA Observatory, NAOJ.

\end{document}